\pdfoutput=1
\documentclass[sigconf]{acmart}

\AtBeginDocument{%
  \providecommand\BibTeX{{%
    \normalfont B\kern-0.5em{\scshape i\kern-0.25em b}\kern-0.8em\TeX}}}

\setcopyright{acmcopyright}
\copyrightyear{2018}
\acmYear{2018}
\acmDOI{XXXXXXX.XXXXXXX}

\acmConference[Submitted to SIGCSE '24]{Submitted to 2024 ACM SIGCSE Technical Symposium}{August 20--23, 2023}{Portland, OR, USA}
\acmPrice{15.00}
\acmISBN{978-1-4503-XXXX-X/18/06}

\newcommand{\universityfullname}{Arizona State University}

\newcommand{\courseyear}{2018}
\newcommand{\coursesemester}{Fall \courseyear}

\newcommand{\cpp}{{\tt C++}}

\usepackage{tikz}
\usepackage{pgfplots}
\pgfplotsset{compat=1.13}

\usepackage{listings}
\begin{document}

\title{Creation of a CS1 Course with Modern C++ Principles}

\author{Ryan E. Dougherty}
\email{ryan.dougherty@westpoint.edu}
\orcid{0000-0003-1739-1127}
\affiliation{%
  \institution{United States Military Academy}
  \city{West Point}
  \state{New York}
  \country{USA}
  \postcode{10996}
}

\renewcommand{\shortauthors}{Ryan E. Dougherty}

\begin{abstract}
Best practices in programming need to be emphasized in a CS1 course as bad student habits persist if not reinforced well.
The \cpp\ programming language, although a relatively old language, has been regularly updated with new versions since 2011, on the pace of once every three years.
Each new version contains important features that make the \cpp\ more complex for backwards compatibility, but often introduce new features to make common use cases simpler to implement.
This poster contains experiences in designing a CS1 course that uses the \cpp\ programming language that incorporates ``modern'' versions of the language from the start, as well as recent conferences about the language. 
Our goals were to prevent many common bad habits among \cpp\ programmers. 
\end{abstract}



\received{20 February 2007}
\received[revised]{12 March 2009}
\received[accepted]{5 June 2009}

\maketitle

\section{Introduction}

According to ConsumerReports\footnote{https://advocacy.consumerreports.org/research/report-future-of-memory-safety/}, between 60 and 70 percent of software vulnerabilities are due to memory unsafety.
It was also gave examples of languages that perpetuate such unsafety, which include \cpp. 
Indeed, the \cpp\ language (among others) is forgiving in letting the programmer do nearly anything, including arbitrary memory accesses.
One of the trade-offs is that programs written in \cpp\ compared to other languages, such as Python, are significantly faster.
Modern versions of the \cpp\ language within the last decade have introduced new features and enhancements that, although do not remove the previous ``unsafe'' features for backwards compatibility, allow for programmers to write less such code.

Since the \cpp\ language will be in use for many years to come, any educators who need to teach this language need to do so effectively, especially within a CS1 setting and/or a course taken by non-majors.
By ``effectively'', we mean in a way that promotes not only safe (in terms of memory) programming practices, but in a style that \cpp\ developers have largely adopted and would understand.
There are several goals that we should consider.
First, students need to be made aware of not only the language itself (and programs within it), but also of these ``unsafe'' features.
Many online tutorials will still contain \cpp\ code with these features, and students may be tempted to follow such examples.
Second, educators need to teach a way of programmatic thinking that easily generalizes to larger and more complex problems.
\cpp\ is a typed language, unlike Python, and thus adapting existing (typed) code in \cpp\ to work in another scenario involves advanced features, such as templates and inheritance. 
Although a full understanding of how these features work may be inaccessible to CS1 students, they should have some knowledge of how to \emph{use} these features appropriately. 
In the context of our institution, the vast majority of students taking the \cpp\ course are non-majors; thus it is vital that we prepare them appropriately as they will likely not be taking another programming course.
In this poster we give our experiences in designing a CS1 course involving ``modern'' \cpp.

\section{Course Decisions}\label{sec:course_context}

We outline the decisions we made in designing a course at \universityfullname\ in \coursesemester\ with modern \cpp\ versions in mind, with a focus on removing any common student bad habits, especially those relating to memory unsafety. 
There are limited instances where ``modern'' features can make an entrance in a CS1 course, but our goal was to introduce the mindset of preferring using these features over historic ones when solving ``big'' problems. 

A common source of frustration for CS1 students is with off-by-one errors, and this source sometimes comes from iterating over some linear data structure by index. 
Consider the simple example in Listing~\ref{lst:simple_iter}, where a raw integer array \lstinline{a} is created, and then iterates over each of the elements in \lstinline{a}.
In \cpp11, the range-based \lstinline{for} loop was introduced, as demonstrated in Listing~\ref{lst:range_for}.
In our course, we sought to not teach indices for as long as possible.

\begin{lstlisting}[caption={Simple Iteration Example},label={lst:simple_iter}]
int a[100];
for (int idx=0; idx < 100; idx++) {
    do_something(a[idx]);
}
\end{lstlisting}

\begin{lstlisting}[caption={Ranged-Based For Loop Example},label={lst:range_for}]
int a[100];
for (int elem : a) {
    do_something(elem);
}
\end{lstlisting}

At this point in the course, we would introduce students to references, which allow modifying data within a collection in-place. 
We give two simple choices: one that involves a constant reference to the original object, and another that is modifiable.
Both of these choices are demonstrated in Listing~\ref{lst:range_for_better}.

\begin{lstlisting}[caption={Range-Based For Loop with References},label={lst:range_for_better}]
int a[100];
// non-modifiable version
for (const int& elem : a) {
    // elem += 1; // does not compile now
    do_something(elem);
}
// modifiable version
for (int& elem : a) {
    elem += 1; // modifies element in place
    do_something(elem);
}
\end{lstlisting}

\cpp\ is a language that allows for dynamic memory allocation on the heap.
Observe Listing~\ref{lst:mem_leak}, which contains an example memory leak.
If the \lstinline{if} statement is taken, then a memory leak occurs.
We then advocate using a managed container, such as \lstinline{std::vector}.

\begin{lstlisting}[caption={Memory Leak},label={lst:mem_leak}]
int func() {
    int *a = new int[n];
    ... // modify a
    if (a[0] > 100) {
        return 10; // missing "delete" here.
    }
    delete[] a;
}
\end{lstlisting}

Unless warnings are turned on, there is another bug that using a managed container solves.
Suppose the student accidentally wrote \lstinline{delete} instead of \lstinline{delete[]}.
We would show students the (optimized) assembly of both versions, proving that the compiler can optimize away the \lstinline{new}/\lstinline{delete[]} version of this code, but not the \lstinline{new}/\lstinline{delete} version.
This has three main benefits for students: (1) there is in fact undefined behavior, (2) there exist different compiler options to help them debug code, and (3) ``warnings \emph{are} errors'' can be enforced via the \lstinline{-Werror} flag. 

Sometimes using a ``raw'' pointer is necessary; for example, legacy code or library functions may require them.
In \cpp11, smart pointers were introduced to automatically take care of this problem; the two types are \lstinline{std::unique_ptr} for unique ownership of an object, and \lstinline{std::shared_ptr} for shared ownership.
The creation of a \lstinline{std::unique_ptr} object with \lstinline{new int} passed to its constructor is not inherently wrong in isolation, \textit{assuming the} \lstinline{std::unique_ptr} \textit{will be fully constructed}.
However, consider a slightly generalized example in Listing~\ref{lst:smart_ptr_func_ex}; it contains a function which takes two \lstinline{std::unique_ptr} objects.
If the programmer is using a version of \cpp\ before \cpp17, there is an unspecified order to the evaluation to the call to \lstinline{some_func}.
Specifically, if the evaluation order is to first dynamically allocate \lstinline{X}, then call that of \lstinline{Y}, which throws an exception (i.e., before the \lstinline{std::unique_ptr} surrounding \lstinline{X} is constructed), then \lstinline{X}'s memory will be leaked as there is no mechanism to \lstinline{delete} it.
Since \cpp17, an argument evaluation order has been standardized.

\begin{lstlisting}[caption={Unique Pointer Function Example},label={lst:smart_ptr_func_ex}]
void some_func(std::unique_ptr<X> T1, std::unique_ptr<Y> T2) { ... }
...
int main() {
    ...
    some_func(std::unique_ptr<X>(new X), std::unique_ptr<Y>(new Y));
}
\end{lstlisting}

In the context of our course design, there are several competing objectives.
On one hand, this is often the first and only programming course students will take.
It would be unwise to give all of the \cpp\ specification details on the behavior of smart pointers and dynamic memory. 
On the other hand, showing and demonstrating code examples that use \lstinline{new} may give students false impressions about how modern \cpp\ code is written.
Additionally, managing this memory (other than in a small example solely to demonstrate memory leaks) will detract from students' focus on the programming task at-hand.

A common source of student programming bugs is with silent type conversions, especially with converting from a ``smaller'' numerical type to a ``larger'' one; we introduce the \lstinline{auto} keyword to showcase how these bugs can be fixed. 
This happens fairly early along in the course with the previously discussed motivations, as well as loop examples with \lstinline{const} reference to avoid copies.

Templates allow for a function, class, etc. to be written effectively once based on one or more (internally held) types, and any variation that use a different internally held type just needs to invoke the template to automatically generate the necessary code. 
In the context of a CS1 course, templates are not as much of a concept students \emph{implement} directly, but rather \emph{understand} why \cpp\ code operates the way it does.
Templates help solve memory unsafety as types (generally) are enforced throughout the function/class.

Functions should not return by \lstinline{const} value, or by reference (most of the time); the former concerns efficiency, and latter concerns memory unsafety (dangling reference).
Of course, other standard coding practices are taught, such as using indentation and curly brackets for ease of reading.
Implied previously, we heavily stress using and understanding compiler flags.
Specifically to \cpp, we advocated not using the line \lstinline{using namespace std;}.

\begin{acks}
The opinions in the work are solely of the author, and do not necessarily reflect those of the U.S. Army, U.S. Army Research Labs, the U.S. Military Academy, or the Department of Defense.
\end{acks}


\end{document}